\documentclass[aip,jmp,nofootinbib]{revtex4-1}
\usepackage{graphics}
\usepackage{graphicx}
\usepackage{dcolumn}
\usepackage{bm}
\usepackage{mathrsfs}
\usepackage{pstricks}
\usepackage{color}
\usepackage{slashed}
\usepackage{amsmath}
\usepackage{epsfig}
\usepackage{amsfonts}
\usepackage{amssymb}
\def\beq{\begin{equation}}
\def\eeq{\end{equation}}
\def\bea{\begin{eqnarray}}
\def\eea{\end{eqnarray}}
\def\nn{\nonumber}

\def\x{\mathbf{x}}

\def\k{\mathbf{k}}

\def\n{\mathbf{n}}
\def\m{\mathbf{m}}
\def\a{\mathbf{a}}
\def\N{\mathbf{N}}
\begin{document}

\title{Light-cone fluctuations and the renormalized stress tensor of a massless scalar field}
\author{V. A. De Lorenci}
\email{delorenci@unifei.edu.br}
\affiliation{Instituto de Ci\^encias, Universidade Federal de Itajub\'a, Itajub\'a, MG 37500-903,
Brazil}
\author{G. Menezes}
\email{gsm@ift.unesp.br}
\affiliation{Instituto de F\'\i sica Te\'orica, Universidade Estadual Paulista,
S\~ao Paulo, SP 01140-070, Brazil}
\author{N. F. Svaiter}
\email{nfuxsvai@cbpf.br}
\affiliation{Centro Brasileiro de Pesquisas F\'{\i}sicas, Rio de Janeiro, RJ 22290-180,
Brazil}

\begin{abstract}
We investigate the effects of light-cone fluctuations over the renormalized vacuum expectation value of the stress-energy tensor of a real massless minimally coupled scalar field defined in a ($d+1$)-dimensional flat space-time with topology ${\cal R}\times {\cal S}^d$. For modeling the influence of light-cone fluctuations over the quantum field, we
consider a random Klein-Gordon equation.
We study the case of centered Gaussian processes.
After taking into account all the
realizations of the random processes, we present the correction caused by random fluctuations. The averaged renormalized vacuum expectation value of the stress-energy associated with the scalar field is presented.
\end{abstract}


\maketitle

\section{Introduction}
\label{intro}

The concept of space-time in special relativity is based on the geometric construction of
fixed light cones, which divides space-time into causally distinct regions. On the other hand, within the general relativity framework, the causal structure of events is not fixed for all times; rather, it is dynamical. Such a geometrical picture of the gravitational field is a very successful classical field theory. There are many attempts to bring general relativity into the quantum domain~\cite{carlip,ash}. Despite such enormous efforts, so far there is no consensus on what would be a quantum theory of gravity. One should expect that one of the consequences of assuming that the gravitational field obeys quantum-mechanical laws is that the structure of space-time must undergo quantum fluctuations. Ford and collaborators developed this idea \cite{ford11,fn2}, showing that the effects of fluctuations of the geometry of the space-time caused by quantum mechanical fluctuations of the gravitational field is to smear out the light cone. In general, in addition to quantum mechanical metric fluctuations, there are induced metric fluctuations generated by quantum fluctuations of matter fields. In both scenarios the concept of light-cone structure has to be modified.

On the other hand, it is well known that when quantum fields are defined in space-times with non-trivial topological structures, interesting effects arise which are connected with vacuum fluctuations of the field. Even if the space-time is unbounded, if the quantum field is constrained by the presence of material boundaries, non-trivial consequences will show up due to vacuum fluctuations. In this respect, the most well known physical manifestation of such fluctuations is the Casimir effect, which has been extensively discussed in the literature~\cite{casimir,plunien,mamayev,bordag,milton,milton2}. In such a context, an intriguing question would be how light-cone fluctuations could affect measurable effects associated with virtual processes in quantum theory. Here we provide a scenario where this issue is investigated.

In this paper we consider the effects of light-cone fluctuations upon the renormalized vacuum expectation value of the stress-energy tensor of a quantum field. In such a direction, recently two of us investigated the influence of light-cone fluctuations over the transition probability rate of a two-level system coupled to a massless scalar field undergoing uniformly accelerated motion~\cite{power}. The assumptions made in such a reference were that the light-cone fluctuations can be treated classically and their effects on the quantum fields can be described via random differential equations.

Here we study the renormalized vacuum expectation value of the stress tensor associated with a real massless minimally coupled scalar field in the presence of a disordered medium. For modeling the influence of light-cone fluctuations over the quantum field, we consider a random Klein-Gordon equation \cite{krein1}. We study the case of centered Gaussian processes. We consider fields defined in a nonsimply connected space-time with topology of ${\cal S}^d \times {\cal R}^1$, with ${\cal S}^d$ corresponding to a hypertorus, i.e., a flat space-time with periodicity in all of the $d$ spatial dimensions. In order to obtain our results we have to implement a perturbation theory associated with a massless scalar field in disordered media
\cite{pe13,pe14,prd}. One could also regard this situation as a simplified model for the more realistic case of the Casimir energy due to phonons~\cite{p2,ford80,p3} in a random fluid confined between two infinite walls.

The organization of the paper is as follows. In section II we present a brief review of the
point-splitting approach that can be used to obtain the renormalized vacuum energy of quantum fields in
the presence of classical macroscopic boundaries and also in curved space-time.
We discuss the modifications in a free scalar quantum field theory due to the presence of randomness in section III.
In section IV we derive the correction caused by random fluctuations in the renormalized vacuum expectation value of the stress tensor associated with quantum fields in the case mentioned above. Conclusions are given in
section V. In this paper we use $32\pi G=\hbar=k_{B}=c=1$.

\section{Renormalized vacuum expectation value of the stress tensor of quantum fields}
\label{ren}

The problem of renormalization of ill defined quantities leading to a physically meaningful result is a fundamental question in
quantum field theory. Although formally divergent, in the absence of gravity the difference between the vacuum energy of quantum fields at different physical configurations can be finite~\cite{casimir}. The formal definition of the Casimir energy is $E_{cas}=E_{0}(\partial\Gamma)-E_{0}(0)$ where $E_{0}(\partial\Gamma)$ is the vacuum energy of a quantum field in
the presence of boundaries and $E_{0}(0)$ is the vacuum energy in free space. For example, the introduction of a pair of conducting
plates into the vacuum of the electromagnetic field alters the zero-point fluctuations of the field and thereby produces an
attraction between the plates \cite{plunien,mamayev,bordag,milton,milton2}. In this paper we are interested to study the effects of light-cone fluctuations over the renormalized vacuum expectation value of the stress tensor.

There are different ways to find the renormalized vacuum energy of quantum fields defined in a flat space-time with nontrivial topology or in generic curved space-times~\cite{green,sinal,blau,nami1,hawking,zeta1,zeta2,fulling,fluct}. Of our particular interest is the point-splitting regularization method. In this approach we compute the vacuum energy from the renormalized stress tensor. For a minimally coupled massless spin-$0$ field defined in flat space-time with a given interaction potential $V(\varphi)$, the stress-tensor reads
\beq
T_{\mu\nu}(x) = \partial_{\mu}\varphi(x)\partial_{\nu}\varphi(x) - \frac{1}{2}\eta_{\mu\nu}\eta^{\alpha\beta}\partial_{\alpha}\varphi(x)\partial_{\beta}\varphi(x)
+ \,\eta_{\mu\nu}\,V(\varphi).
\eeq
The potential $V(\varphi)$ may depend on products of fields and field derivatives.~Here $\eta_{\mu\nu}$ is the usual Minkowski metric (we take the sign convention of \cite{davis}). For a quantum field propagating in a medium with disorder, $V(\varphi)$ will represent the coupling between the field and random impurities.~In this paper, we assume that such a random potential has the following functional form:
\beq
V(\varphi) = \frac{1}{2}\,\nu(x)\partial_0\varphi(x)\partial_0\varphi(x).
\label{rp}
\eeq
The statistical properties of the random variable $\nu(x)$ which describes randomness will be given in due course. Hence, using the point-splitting method, the vacuum expectation value of the stress tensor is found to be
\begin{equation}
\langle T_{\mu\nu}(x) \rangle = \lim_{x',x''\rightarrow x}{\cal T}_{\mu\nu}(x'',x')
\label{tensor21}
\end{equation}
where
\bea
{\cal T}_{\mu\nu}(x'',x') &=& \frac{1}{4}\biggl[2\,\partial_{\mu''}\partial_{\nu'}G^{(1)}(x'',x')
-\,\eta_{\mu\nu}\eta^{\alpha\beta}\partial_{\alpha''}\partial_{\beta'}G^{(1)}(x'',x')
\nn\\
&& +\, \eta_{\mu\nu}\,\nu(x)\partial_{0''}\partial_{0'}G^{(1)}(x'',x')\biggr],
\label{tensor22}
\eea
where the Green's function $G^{(1)}(x,x')$ is given by
\begin{equation}
G^{(1)}(x,x') = \langle \{\varphi(x),\varphi(x')\} \rangle = G^{(+)}(x,x') + G^{(-)}(x,x'),
\label{hadam}
\end{equation}
with the Wightman functions given by the expressions $G^{(+)}(x,x') = \langle \varphi(x)\varphi(x')
\rangle$ and $G^{(-)}(x,x') = \langle \varphi(x')\varphi(x) \rangle$. In Eq.~(\ref{tensor22}) it is to be understood that $\partial_{\mu'}$ ($\partial_{\mu''}$) acts on $x'$ ($x''$). We chose to write such an equation in a form which is symmetric under the interchange $x''\leftrightarrow x'$. We remark that we are using the summation convention, i.e., repeated indices are summed unless otherwise stated. Greek indices are referred to space-time components (e.g., $\mu, \nu, \rho, \cdots = 0, 1, \cdots, d$) whereas latin indices stand for space components (e.g., $i, j, r,\cdots = 1, 2, \cdots, d$). 

The coincidence limits for $G^{(1)}(x,x')$ and its derivatives yield formally divergent expressions. Therefore Eq.~(\ref{tensor21}) should be properly renormalized. Considering a flat background space-time, this could be done in the usual way, i.e., by replacing such a quantity by the renormalized expectation value of the stress tensor which is given by
\begin{equation}
\langle: T_{\mu\nu}(x) :\rangle = \langle T_{\mu\nu}(x) \rangle - \langle T_{\mu\nu}(x) \rangle_M,
\label{tensor23}
\end{equation}
where $\langle T_{\mu\nu}(x) \rangle_M$ is the expectation value of the stress tensor in Minkowski space-time. In this way, the renormalized vacuum energy density is given by $\langle: T_{00}(x) :\rangle$. We remark that such a procedure cannot be trusted when the background space-time is curved. In non-gravitational physics, only energy differences are observable; in this case the subtraction scheme can be carried out without further inconveniences. When one considers gravity, this technique is not satisfactory since energy is a source of gravity and therefore we are not free to reescale the zero-point energy.

In the next section we discuss a scalar quantum field theory in the presence of stochastic fluctuations of the light cone.

\section{Scalar quantum field theory in the presence of random fluctuations}
\label{flu}

In this section, we present the solution to the scalar field equation in the
presence of light-cone fluctuations. As pointed out, for modeling the influence
of such fluctuations over the quantum field, it is enough
to consider a stochastic Klein-Gordon equation. The field equation obtained in this
way cannot be solved in a closed form. However, assuming that stochastic fluctuations
are small, one may introduce a perturbation theory similar to the one discussed
in~\cite{prd}. So, we will have an expression for the Hadamard function that will
contain the corrections due to light-cone fluctuations. This will enable us to
calculate the vacuum expectation value of the stress tensor and, therefore, the
corrections to the Casimir energy.

Before we describe the physical situation which will lead us to the vacuum energy,
let us present the field equation for a massless minimally coupled scalar field
$\varphi$ in a space-time with stochastic light-cone fluctuations. It reads~\cite{prd}
\begin{equation}
\left\{\left[1 + \nu(x)\right]\frac{\partial^{2}}{\partial
t^{2}}-\nabla^{2}\right\}\varphi(x) = 0.
\label{randomeq}
\end{equation}
We remark that such an equation can be derived from a Lagragian containing the random potential given by Eq.~(\ref{rp}), assuming that the inhomogeneity and disorder are smooth, i.e., $\nu(x)$ is a slowly-varying function in comparison with $\varphi$. In this way, derivatives of the $\nu(x)$ are neglected. The important point to be observed is that we are not considering a toy model for quantum gravity; rather, we are interested in how quantum fields behave in the presence of random fluctuations of the light cone.

For simplicity, we consider the (local) random variable $\nu(x)$ to be a Gaussian centered distribution:
\begin{equation}
\overline{\nu(x)} = 0,
\label{nami21}
\end{equation}
with a white-noise correlation function given by:
\begin{equation}
\overline{\nu(x)\nu(x')} =
\sigma^2\,\delta(x-x'),
\label{nami22}
\end{equation}
where $\sigma^2$ gives the intensity of stochastic fluctuations and $\delta(x-x')$
is the $(d+1)$-dimensional Dirac delta function.
The symbol $\overline{(\cdots)}$ denotes an average over
all possible realization of the random variable.
On the other hand, in principle it is possible to extend the method to
colored or non-Gaussian noise functions. In addition, observe that
the noise sources define a preferred reference frame, similarly
to an external heat bath, which induces a breaking in Lorentz symmetry.

The field equation (\ref{randomeq}) should be compared with the ones used
in Refs. \cite{krein1,prd}. However, in contrast to these references, which
employ a static noise, here we assume that the random function is also
time-dependent. This situation was carefully analyzed in Ref.~\cite{time}. Since the solution to the field equation cannot
be given in a closed form, one can employ a perturbative series expansion for Green's functions. The propagator $iG(x,x')=\langle\,T\,(\varphi(x)\varphi(x'))\rangle$ associated with the wave equation~(\ref{randomeq}) satisfies
\begin{equation}
\left[\left[1 + \nu(x)\right]\frac{\partial^{2}}{\partial
t^{2}}-\nabla_{\x}^{2}\right] G(x,x')= -\delta(t-t')\delta^{d-1}(\x-\x').
\label{p14}
\end{equation}
Suitable boundary conditions must be imposed on the solutions of the above equation in order for them to have the properties of a time-ordered product. In order to accommodate the appropriate modifications introduced by the presence of the random term, we proceed as follows. Define the operator $K = K_0 - L$, where $K_0(x,y) = \Bigl(\partial^{2}/\partial t^{2}-\nabla_{\x}^{2}-i\epsilon\Bigr)\delta(x-y)$ and
\beq
L(x) = - \nu(x) \frac{\partial^2}{\partial t^2}.
\label{L1-coord}
\eeq
This leads to the following formal relation:
\begin{equation}
[K(x,y)]^{-1} = - G(x,y).
\end{equation}
In this way, since the disorder is ``weak", a natural perturbative expansion for $G$ in form of a Dyson series can be defined:
\begin{equation}
G = G_0 - G_0\,L\,G_0 + G_0\,L\,G_0\,L\,G_0 + \cdots
\label{dy}
\end{equation}
where we used a formal operator notation in this last equation and $[K_0]^{-1}
= - G_0$ is the unperturbed propagator. Following Ref.~\cite{prd},
equation (\ref{dy}) can be written in terms of space and time variables:
\beq
\hspace{-0.25cm}
G(x,x') = G_{0}(x,x') +
\sum_{n=1}^{\infty}\int dy_1 \,G_{0}(x,y_1){\cal G}^{(n)}(y_1,x'),
\label{a1}
\eeq
where
\begin{equation}
{\cal G}^{(n)}(y_1,x') = (-1)^{n}\prod_{j=1}^n L(y_j)
\int dy_{j+1} \,G_{0}(y_j,y_{j+1}).
\label{genericterm}
\end{equation}
In Eq.~(\ref{genericterm}), it is to be understood that $y_{n+1} = x'$ and
that there is no integration in $y_{n+1}$.
Details on the derivations of the above expressions can be found in
Ref.~\cite{prd}. Furthermore, due to the Gaussian nature of the noise
averaging, higher order correlation functions
of the form $\overline{\nu(x_1)\nu(x_2)\cdots\nu(x_p)}$
can be easily expressed as the sum of products of two-point correlation functions
corresponding to all possible partitions of ${x}_{1},{x}_{2}, \cdots , {x}_{p}$.

For the purposes in the present paper we consider terms up to second order
in $\nu$ of the above series. The corrections to $G$ are discussed at length in the appendix. With an expression for
the propagator, one is able to calculate the Green's function $G^{(1)}(x,x')$ through
the following formulae. The propagator can be written
in terms of the Wightman functions as $iG(x,x') = \theta(t-t')G^{(+)}(x,x')
+ \theta(t'-t)G^{(-)}(x,x')$. In turn, since $G^{(1)}(x,x')$ is given by Eq.~(\ref{hadam}), one sees that the calculations of the propagator allows one to reach expressions for the Green's function $G^{(1)}(x,x')$ in situations where one can use the decomposition property stated above. On the other hand, $G^{(1)}(x,x')$ can also be computed through the relation~\cite{davis}
\begin{equation}
G(x,x')  + \frac{1}{2}\Bigl[G^R(x,x') + G^A(x,x')\Bigr] = -\frac{i}{2}\,G^{(1)}(x,x'),
\label{apg2}
\end{equation}
where $G^{R}(x,x') = i\theta(t-t')\langle [\varphi(x),\varphi(x')] \rangle$ ($G^{A}(x,x') = -i\theta(t'-t)\langle [\varphi(x),\varphi(x')] \rangle$) is the retarded (advanced) Green's function which obeys $[K]^{-1} =  G^{R(A)}$.

After this digression on Green's functions, we are able to calculate the corrections to the expectation value of the vacuum energy due to randomness of the light cone. This is the subject of the next section.

\section{The averaged renormalized vacuum expectation value of the stress tensor}
\label{stress-tensor}

The aim of this section is to determine the averaged components of the renormalized vacuum expectation value of the stress tensor. Such quantities will be calculated in the way prescribed in Sec.~\ref{ren}. As mentiond above, the vacuum energy density is given by $\langle T_{00}(x) \rangle$. Other stress-tensor components have well known physical interpretation. We should average such quantities over all the realizations of the noise. Therefore, after performing the stochastic averages of Eq.~(\ref{tensor22}), the vacuum expectation values of the stress-tensor components are given by the coincidence limit of the following expression
\bea
\overline{{\cal T}_{\mu\nu}(x'',x')} &=& \frac{1}{4}\biggl\{2\,\partial_{\mu''}\partial_{\nu'}\overline{G^{(1)}(x'',x')}
- \eta_{\mu\nu}\eta^{\alpha\beta}\partial_{\alpha''}\partial_{\beta'}\overline{G^{(1)}(x'',x')}
\nn\\ &&
+ \eta_{\mu\nu}\,\partial_{0''}\partial_{0'}\overline{\nu(x)\,G^{(1)}(x'',x')}\biggr\}.
\label{e}
\eea
Now we focus our attentions on calculating the components of the stress tensor considering the effects of light-cone random fluctuations. We consider the case where the fields satisfy periodic boundary conditions in all spatial directions. In this way, since the background space-time is flat, we may use the subtraction scheme discussed in Sec.~\ref{ren}. The corrections to the Green's function $G^{(1)}(x,x')$ up to second order in the perturbations are given in the Appendix. Let us first present the zero-order contribution. Inserting expression (\ref{gns}) in Eq.~(\ref{e}) and remembering Eqs.~(\ref{tensor21}) and~(\ref{tensor23}) leads to the following expression for the renormalized vacuum energy density
\begin{equation}
\langle:T_{00}(x):\rangle_{0} = \frac{1}{2a_1}\sum_{n_1 = -\infty}^{+\infty}\cdots\frac{1}{a_d}\sum_{n_d = -\infty}^{+\infty}\,k(\n),
\label{normal}
\end{equation}
where the $n = 0$ term is excluded from the above sum as it is just the contribution from the Minkowski vacuum. Following the discussion presented in the Appendix, we write this multiple sum in terms of
the Epstein zeta function. Using~(\ref{zetep}) and defining
\beq
f(d) = \frac{\Gamma\bigl(
\frac{d+1}{2}\bigr)}{2\pi^{(d+1)/2}}
\label{efe}
\eeq
the renormalized vacuum energy density becomes
\begin{equation}
\langle:T_{00}(x):\rangle_{0} = -\,f(d)\,Z(a_1,...,a_d;d+1).
\label{eosp}
\end{equation}
The result~(\ref{eosp}) is finite for all $d > 0$ and is always negative. To reach such an expression we must remember to introduce an arbitrary mass parameter $\mu$ in the summations above in order to keep the Epstein zeta funtion a dimensionless quantity. This procedure is necessary in order to enable one to implement the analytic procedure described in the Appendix.

Similarly:
\bea
&\langle: T_{ij}(x) :\rangle_{0}&\,= \frac{1}{4a_1}\sum_{n_1 = -\infty}^{+\infty}\cdots\frac{1}{a_d}\sum_{n_d = -\infty}^{+\infty}\,\frac{1}{k(\n)}
\nn\\ &&
\,\times\biggl[2\,k_i k_j - \eta_{ij}\eta^{rs}k_r k_s -\eta_{ij}\,k^2(\n)\biggr],
\eea
where $k_{j} = 2\pi n_{j}a_{j}^{-1}$. For $i = j$, one has
\begin{eqnarray}
\langle: T_{jj}(x) :\rangle_{0} &=& \frac{1}{2a_1}\sum_{n_1 = -\infty}^{+\infty}\cdots\frac{1}{a_d}\sum_{n_d = -\infty}^{+\infty}\,\frac{k_{j}^2}{k(\n)}
\nonumber \\
&=&  \frac{1}{2\,d\,a_1}\sum_{n_1 = -\infty}^{+\infty}\cdots\frac{1}{a_d}\sum_{n_d = -\infty}^{+\infty}\,k(\n),
\end{eqnarray}
where the last equality follows by symmetry. In the above equation it is to be understood that there is no summation over the repeated index $j$. Proceeding as above, one has:
\begin{equation}
\langle: T_{jj}(x) :\rangle_{0} = -\,\frac{f(d)}{d}\,Z(a_1,...,a_d;d+1).
\label{posp}
\end{equation}
Now consider $i \neq j$. One has
\beq
\langle: T_{ij}(x) :\rangle_{0} = \frac{1}{2a_1}\sum_{n_1 = -\infty}^{+\infty}\cdots\frac{1}{a_d}\sum_{n_d = -\infty}^{+\infty}\,\frac{k_i\,k_j}{k(\n)}.
\eeq
Since we have sums over the integers of a product between an even function and an odd function, we have that $\langle: T_{ij}(x) :\rangle_{0} = 0$ for $i \neq j$.

As for the zero-order contribution to momentum density, one has, after considering the coincidence limit and using~(\ref{tensor23})
\beq
\langle: T_{0j}(x) :\rangle_{0} = - \frac{1}{2a_1}\sum_{n_1 = -\infty}^{+\infty}\cdots\frac{1}{a_d}\sum_{n_d = -\infty}^{+\infty}\,k_j
\eeq
The sums over indices other than $j$ in the above expression can be expressed with the help of a integral representation for the Epstein zeta-function. It gives an analytic continuation for such series except for a pole at $p=2s$~\cite{ambjorn}. It is
%
\bea
&&(\pi\,\eta)^{-s}\,\Gamma(s)\,Z_p\,(a_1,\cdots,a_p\,;2s) = -\frac{1}{s}+\frac{2}{p-2s}
\nn\\&&
+\eta^{-s}\int_{\eta}^{\infty}dx\,x^{s-1}
\left(\vartheta(0,\cdots,0;a_1^2\,x,\cdots,a_p^2\,x)-1\right)
\nn\\
&&+\eta^{\frac{2s-p}{2}}\int_{1/\eta}^{\infty}dx\,x^{(p-2s)/2-1}
\,\Biggl(\vartheta\left(0,\cdots,0;\frac{x}{a_1^2},\cdots,\frac{x}{a_p^2}\right)-1\Biggr)\,,
\label{epstein2}
\eea
%
where $\eta^{\,p/2}$ is the product of the $p\,'s$ parameters
$a_{i}$ given by $\eta^{\,p/2}=a_1 \cdots a_p$, and the generalized
Jacobi function $\vartheta(z_1,\cdots,z_p;x_1,\cdots,x_p)$, is defined
by
\begin{equation} \vartheta(z_1,...,z_p\,;x_1,...,x_p)=\prod_{i=1}^{p}\vartheta(z_i;x_i)\,,
\end{equation}
with $\vartheta(z;x)$ being the Jacobi function, i.e.,
\begin{equation}
\vartheta(z;x)=\sum_{n=-\infty}^{\infty}e^{\pi(2nz-n^2x)}\,.
\end{equation}
Using this integral expression for the Epstein zeta-function,
given by Eq.~(\ref{epstein2}), we can find that
\begin{equation}
Z_p(a_1,...,a_p\,;2s)|_{s=0}=-1\,,
\label{epstein3}
\end{equation}
for any $p\geq 1$. So
\beq
\langle: T_{0j}(x) :\rangle_{0} = \frac{1}{2 a_1 \cdots a_d}\sum_{n_j = -\infty}^{+\infty}\,k_j.
\eeq
Such a summation is zero, as the reader can easily check. So $\langle: T_{0j}(x) :\rangle_{0} = 0$.

Now let us introduce the corrections due to light-cone random fluctuations. First consider the corrections to the renormalized vacuum energy density. Inserting  Eqs.~(\ref{g111p}) and~(\ref{h1mp}) in Eq.~(\ref{e}) and taking the coincidence limit for the component  $\overline{{\cal T}_{00}(x'',x')}$ yields
\beq
\overline{\langle: T_{00}(x) :\rangle}_{1} = P + E,
\eeq
where
\beq
P = -\frac{\sigma^2}{8 a_1}\sum_{n_1 = -\infty}^{+\infty}\cdots\frac{1}{a_d}\sum_{n_d = -\infty}^{+\infty}1
\frac{1}{a_1}\sum_{m_1 = -\infty}^{+\infty}\cdots\frac{1}{a_d}\sum_{m_d = -\infty}^{+\infty}k^2(\m),
\eeq
and
\bea
E = \sigma^2\,Q(a_1,...,a_d;d)\frac{1}{2 a_1}\sum_{n_1 = -\infty}^{+\infty}\cdots\frac{1}{a_d}\sum_{n_d = -\infty}^{+\infty}k(\n).
\label{strange}
\eea
Consider the quantity $P$. The sums over $n$ can be evaluated considering that $$\sum_{n_1 = -\infty}^{+\infty}\cdots\sum_{n_d = -\infty}^{+\infty}1 = Z_d(a_1,...,a_d\,;2s)|_{s=0}.$$ So, with the help of Eq.~(\ref{epstein3}), we get
\beq
P = \frac{\sigma^2}{8 a_1}\sum_{m_1 = -\infty}^{+\infty}\cdots\frac{1}{a_d}\sum_{m_d = -\infty}^{+\infty}k^2(\m).
\eeq
The sum over $m$ can also be expressed in terms of the Epstein zeta-function $$\sum_{m_1 = -\infty}^{+\infty}\cdots\sum_{m_d = -\infty}^{+\infty}k^2(\m) = (2\pi)^2\,Z_d(a_1,...,a_d\,;-2),$$ which vanishes, in virtue of the functional reflection formula~(\ref{zetep}). So $P = 0$. Consider now $E$. Comparing Eqs.~(\ref{normal}) and~(\ref{strange}), one has
\bea
E = - \sigma^2\,\left[f(d)\,Z(a_1,...,a_d;d+1)\right]^2,
\eea
where we have used Eqs.~(\ref{efe}) and~(\ref{qe}). Therefore, collecting our results one has
\beq
\overline{\langle: T_{00}(x) :\rangle}_{1}\, = - \sigma^2\,\biggl[f(d)\,Z(a_1,...,a_d;d+1)\biggr]^2.
\label{e1sp}
\eeq
This is the correction to the renormalized vacuum energy due to light-cone fluctuations up to second order in the noise. The subscript $``1"$ in the left-hand side of the above equation indicates the first order correction after performing the random averages. Now let us calculate the corrections to the components $\overline{\langle: T_{ij}(x) :\rangle}$. Inserting Eqs.~(\ref{g111p}) and~(\ref{h1mp}) in Eq.~(\ref{e}), one has
\bea
&&\overline{\langle: T_{ij}(x) :\rangle}_{1} = \frac{\sigma^2}{8}\,Q\, \frac{1}{ a_1}\sum_{n_1 = -\infty}^{+\infty}\cdots\frac{1}{a_d}\sum_{n_d = -\infty}^{+\infty}\frac{1}{k(\n)}
\nn\\ &&
\times\biggl[2\,k_i k_j - \eta_{ij}\eta^{rs}k_r k_s
-3\eta_{ij}\,k^2(\n)\biggr] + \eta_{ij}\,P.
\eea
Considering the results derived above one has, for $i \neq j$, $\overline{\langle: T_{ij}(x) :\rangle}_{1} = 0$. For $i = j$, one has (no summation over the index $j$):
\bea
\overline{\langle: T_{jj}(x) :\rangle}_{1} &=& \frac{\sigma^2}{4}\frac{Q}{ a_1}\sum_{n_1 = -\infty}^{+\infty}\cdots\frac{1}{a_d}\sum_{n_d = -\infty}^{+\infty}\frac{k^2_j + k^2(\n)}{k(\n)}
\nn\\ &&
= \frac{\sigma^2}{4}\frac{(d+1) Q}{d}\frac{1}{ a_1}\sum_{n_1 = -\infty}^{+\infty}\cdots\frac{1}{a_d}\sum_{n_d = -\infty}^{+\infty}k(\n),
\eea
where the last line of the right-hand side of the above expression follows by symmetry. Hence, using the same technique as above one has
\beq
\overline{\langle: T_{jj}(x) :\rangle}_{1} = -\frac{\sigma^2}{2}
\left(1+ \frac{1}{d}\right)\,\biggl[f(d)\,Z(a_1,...,a_d;d+1)\biggr]^2.
\label{p1sp}
\eeq
With similar considerations as before, one may show that all corrections to the momentum density vanish, $\overline{\langle: T_{0j}(x) :\rangle}_{1} = 0$. Consequently, the final form for the renormalized expectation value of stress-tensor components in a non-simply connected space-time subjected to light-cone fluctuations reads, up to second order in the perturbations:
\begin{equation}
\overline{\langle: T_{00}(x) :\rangle} = -g(\a,d)\left[1+ \sigma^2\,g(\a,d)\right],
\label{efsp}
\end{equation}
and
\begin{equation}
\overline{\langle: T_{jj}(x) :\rangle} = -g(\a,d)\biggl[\frac{1}{d}+\frac{\sigma^2}{2}
\left(1+ \frac{1}{d}\right)g(\a,d)\biggr],
\label{pfsp}
\end{equation}
where $\a = (a_1,a_2,\cdots,a_d)$ and
\begin{equation}
g(\a,d) = f(d)\,Z(a_1,...,a_d;d+1),
\end{equation}
The above expressions summarize the main results of the paper.  Let us now discuss the results presented here.\\

\section{Discussions and conclusions}
\label{conclude}

In this paper we studied a massless scalar field theory in the presence of light-cone fluctuations. After performing the random averages over the noise function, the correction caused by randomness in the renormalized stress tensor associated with the quantum field was presented. We obtained a correction which is proportional to the square of the unperturbed contribution.

We remark that, although the Casimir effect is a well understood phenomenon, there are still some open questions related to this effect. A interesting question is how the sign of the Casimir force depends on the topology, dimensionality of the space-time, the shape of bounding geometry or others physical properties of the system \cite{ambjorn,namis}. This problem is still unsolved in the literature. There are also some controversies in the literature that inspired many recent papers. For example, questions concerning the temperature dependence of real materials and also how to obtain closed-form results for the interaction of bodies that alters the zero-point energy of the electromagnetic field. Here we did not consider such problems.

A natural extension of this paper is to study the renormalized vacuum energy due to phonons in a disordered fluid confined between plane boundaries. Phonons share several properties with relativistic quantum fields. Quantized acoustic perturbations in the presence of disorder and boundaries lead us to the phononic Casimir effect with randomness. This subject is under investigation by the authors.


\section{Acknowlegements}

We would like to thank E. Arias for useful discussions.
This paper was supported by the Brazilian agencies CAPES, CNPq and FAPEMIG.

\appendix
	
\section{Perturbative corrections to the Feynman propagator and the Hadamard function for a nonsimply connected space-time}
\label{app2}

Our aim is to present an expression for the Hadamard function from which we can calculate corrections to $\langle T_{\mu\nu} \rangle$ due to light-cone fluctuations. Let us present the propagator up to second order in $\nu(x)$. From (\ref{a1}), we have:
\begin{eqnarray}
G(x,x') &=& G_{0}(x,x') - \int\,dy\,G_{0}(x,y)L(y)G_{0}(y,x')   \nonumber\\ &&
+ \int\int\,dy_1\,dy_2\,G_{0}(x,y_1)L(y_1)G_{0}(y_1,y_2)L(y_2)G_{0}(y_2,x').
\label{exp}
\end{eqnarray}
We consider a topology of the background space-time such that the fields must satisfy
periodic boundary conditions in all spatial directions. For a general hypercuboidal space, with sides of finite length $a_1,...,a_d$, this corresponds to the compactification of the space dimensions to a hypertorus ${\cal S}^d$. The modes of the field then consist of a simple product of modes analogous to the usual Minkowski space. In this way, the unperturbed propagator reads
\begin{eqnarray}
G_{0}(x,x') &=& \frac{1}{a_1}\sum_{n_1 = -\infty}^{+\infty}\!\!\cdots\frac{1}{a_d}
\sum_{n_d = -\infty}^{+\infty}\int\frac{d\omega}{(2\pi)} \,
\, \frac{e^{i[\k(\n)\cdot (\x-\x')_{\bot}-\omega(t-t')]} }{\omega^2 -  k^2(\n)+ i\epsilon},
\label{ft1}
\end{eqnarray}
with
\beq
\k(\n) = 2\pi\,\N = 2\pi (n_1/a_1,n_2/a_2,\cdots,n_d/a_d),
\eeq
and
\beq
\k^2(\n) = k^2(\n) = k_1^2 + k_2^2 + \cdots + k_d^2 = \biggl(\frac{2\pi n_1}{a_1}\biggr)^2 + \biggl(\frac{2\pi n_2}{a_2}\biggr)^2 + \cdots + \biggl(\frac{2\pi n_d}{a_d}\biggr)^2,
\eeq
where use has been made of the notation $\n^2 = n_1^2 + n_2^2 + \cdots + n_d^2$. In order to perform the $\omega$ integration we may resort to contour integrals. We choose the usual contour for the Feynman propagator. See for instance~\cite{davis}. We get
\begin{eqnarray}
&G_{0}(x,x')&\, = \frac{-i}{a_1}\sum_{n_1 = -\infty}^{+\infty}\!\!\cdots\frac{1}{a_d}
\sum_{n_d = -\infty}^{+\infty}
\!\frac{1}{2\,k(\n)}\biggl[e^{i[\k(\n)\cdot(\x-\x') - k(\n)(t-t')]}\theta(t-t')
\nonumber\\ &&
+\, e^{-i[\k(\n)\cdot(\x-\x') - k(\n)(t-t')]}\theta(t'-t)\biggr].
\label{gf0p}
\end{eqnarray}
Now let us introduce the corrections due to light-cone random fluctuations. In virtue of Eqs. (\ref{nami21}) and (\ref{nami22}), the first correction to $G$ will be given by the third term on the right-hand side of (\ref{exp}). Then
\begin{equation}
G_{2}(x,x') = \int\int\,dy_1\,dy_2\,G_{0}(x,y_1)L(y_1)G_{0}(y_1,y_2)L(y_2)G_{0}(y_2,x'),
\label{correc}
\end{equation}
where the subscript in $G$ stands for $n$th-order in $\nu(x)$. Inserting Eqs. (\ref{L1-coord}) and (\ref{ft1}) in the last expression and then using (\ref{nami22}) allow us to write:
\bea
\overline{G_{2}(x,x')} &=& \frac{1}{a_1}\sum_{n_1 = -\infty}^{+\infty}
\cdots\frac{1}{a_d}\sum_{n_d = -\infty}^{+\infty}\int\frac{d\omega}{2\pi}
\,e^{i[\k(\n)\cdot (\x-\x')-\omega(t-t')]} \,
\nonumber\\ &&
\times\frac{1}{(\omega^2-{\bf k}^2(\n)+ i\epsilon)}\,\Sigma(\omega)\,
\frac{1}{(\omega^2-{\bf k}^2(\n)+ i\epsilon)},
\label{gf1p}
\eea
where
\begin{equation}
\Sigma(\omega) = \lim_{\delta\rightarrow 0}\sigma^2\omega^2\,\frac{1}{a_1}\sum_{n_1 = -\infty}^{+\infty}\cdots\frac{1}{a_d}\sum_{n_d = -\infty}^{+\infty}\int\frac{d\omega'}{2\pi} \, \frac{\omega'^2\,e^{i\delta\omega}}{\omega'^2-{\bf k}^2(\n) + i\epsilon}.
\label{sigp}
\end{equation}
Performing the $\omega$ integration as above we get, after taking $\delta\rightarrow 0$:
\begin{equation}
\Sigma(\omega) = -\frac{i\sigma^2\omega^2}{2}\frac{1}{a_1}\sum_{n_1 = -\infty}^{+\infty}\cdots\frac{1}{a_d}\sum_{n_d = -\infty}^{+\infty}\,k(\n).
\end{equation}
Such a multiple sum may be written in terms of the Epstein zeta function~\cite{zeta1}:
\begin{equation}
Z(1/a_1,...,1/a_p;s) = \sum_{n_1 = -\infty}^{+\infty}\cdots\sum_{n_p = -\infty}^{+\infty}\,\biggl[\biggl(\frac{ n_1}{a_1}\biggr)^2 + \biggl(\frac{n_2}{a_2}\biggr)^2 + \cdots + \biggl(\frac{n_p}{a_p}\biggr)^2\biggr]^{-s/2},
\label{ep-zeta}
\end{equation}
for $s>p$ and it should be understood that the term for which all $n_i=0$ is to be ommited. This function obeys the reflection formula~\cite{ambjorn}
\begin{equation}
\Gamma\biggl(\frac{s}{2}\biggr)\,\pi^{-s/2}Z(a_1,...,a_p;s)= a_1^{-1}\cdots\,a_p^{-1}\Gamma\biggl(\frac{p-s}{2}\biggr)\,\pi^{(s-p)/2}Z(1/a_1,...,1/a_p;p-s).
\label{zetep}
\end{equation}
Using Eqs.~(\ref{ep-zeta}) and~(\ref{zetep}), we have
\begin{equation}
\Sigma(\omega) =  i\sigma^2\omega^2 Z(a_1,...,a_d;d+1)\frac{\Gamma\bigl(\frac{d+1}{2}\bigr)}{2\pi^{(d+1)/2}}.
\label{sigp1}
\end{equation}
Then, after performing the $\omega$ integration:
\begin{eqnarray}
\overline{G_{2}(x,x')} &=&  \frac{\sigma^2}{2}\,Q(a_1,...,a_d;d)\, \frac{1}{ a_1}
\sum_{n_1 = -\infty}^{+\infty}\cdots\frac{1}{a_d}\sum_{n_d = -\infty}^{+\infty}
\nonumber \\ &&
\times\Biggl[\biggl(\,\frac{1}{2k(\n)} - \frac{i(t-t')}{2}\biggr)e^{i[\k(\n)\cdot(\x-\x')
- k(\n)(t-t')]}\theta(t-t')
\nonumber\\
&&+ \biggl(\,\frac{1}{2k(\n)} -
\frac{i(t'-t)}{2}\biggr)e^{-i[\k(\n)\cdot(\x-\x') - k(\n)(t-t')]}\theta(t'-t)\Biggr],
\label{gf11p}
\end{eqnarray}
where
\begin{equation}
Q(a_1,...,a_d;d) = Z(a_1,...,a_d;d+1)\frac{\Gamma\bigl(\frac{d+1}{2}\bigr)}{2\pi^{(d+1)/2}}.
\label{qe}
\end{equation}
Employing the decomposition property of the propagator in terms of the Wightman functions, one has $$iG_{0}(x,x') + \overline{G_{2}(x,x')} = \langle\,T\,(\varphi(x)\varphi(x'))\rangle = \theta(t-t')\overline{G^{(+)}(x,x')} + \theta(t'-t)\overline{G^{(-)}(x,x')}.$$  Hence, employing Eq.~(\ref{hadam}), one sees that
\begin{eqnarray}
G^{(1)}(x,x') = G^{(1)}_0(x,x') + \overline{G^{(1)}_{2}(x,x')},
\label{g11}
\end{eqnarray}
with the unperturbed Green's function $G^{(1)}(x,x')$ given by
\begin{equation}
G^{(1)}_0(x,x') = \frac{1}{a_1}\sum_{n_1 = -\infty}^{+\infty}\cdots\frac{1}{a_d}\sum_{n_d = -\infty}^{+\infty}\frac{\cos{[\k(\n)\cdot(\x-\x') - k(\n)(t-t')] }}{k(\n)}.
\label{gns}
\end{equation}
Therefore the respective correction to $G^{(1)}(x,x')$ is
\begin{eqnarray}
\overline{G^{(1)}_{2}(x,x')} && = \frac{\sigma^2}{2}\,Q(a_1,...,a_d;d)\, \frac{1}{ a_1}\sum_{n_1 = -\infty}^{+\infty}\cdots\frac{1}{a_d}\sum_{n_d = -\infty}^{+\infty}\Biggl\{\frac{\cos{\bigl[\k(\n)\cdot(\x-\x') - k(\n)(t-t')\bigr] }}{k(\n)}  \nonumber \\ && + (t-t')\sin{\bigl[\k(\n)\cdot(\x-\x') - k(\n)(t-t')\bigr] }\Biggr\}.
\label{g111p}
\end{eqnarray}

Note that, from Eq.~(\ref{e}), the second term on the right-hand side of Eq.~(\ref{exp}) should give contributions to the renormalized stress tensor. Inserting Eq.~(\ref{L1-coord}) in Eq.~(\ref{exp}) and using the Fourier representation~(\ref{ft1}) as well as the noise correlation~(\ref{nami22}), we have (keeping terms up to second order in the noise field):
\begin{eqnarray}
&\overline{\nu(x)\,G_{1}(x'',x')}&\, = -\sigma^2\frac{1}{a_1}\sum_{n_1 = -\infty}^{+\infty}\!\!\cdots\frac{1}{a_d}
\sum_{n_d = -\infty}^{+\infty}\int\frac{d\omega'}{(2\pi)}\,\frac{e^{i[\k(\n)\cdot(\x''-\x)_{\bot} - \omega'(t''-t)]}}{\omega'^2-k^2(\n) + i\epsilon}\nonumber \\ && \times\frac{1}{a_1}\sum_{m_1 = -\infty}^{+\infty}\!\!\cdots\frac{1}{a_d}
\sum_{m_d = -\infty}^{+\infty}\int\frac{d\omega}{(2\pi)} \frac{e^{i[\k(\m)\cdot(\x-\x')_{\bot} - \omega(t-t')]}}{\omega^2-k^2(\m) + i\epsilon}\,\omega^2.
\label{hfm}
\end{eqnarray}
We see that we get a sort of product between two unperturbed propagators. Performing the integration over $\omega$ yields:
\begin{eqnarray}
\overline{\nu(x)\,G_{1}(x'',x')} &=& \frac{\sigma^2}{a_1}\sum_{n_1 = -\infty}^{+\infty}\cdots\frac{1}{a_d}\sum_{n_d = -\infty}^{+\infty}\frac{1}{2k(\n)}\biggl[e^{i[\k(\n)\cdot(\x''-\x) - k(\n)(t''-t)]}\theta(t''-t)  \nonumber \\ && + e^{-i[\k(\n)\cdot(\x''-\x) - k(\n)(t''-t)]}\theta(t-t'')\biggr]\nonumber\\ && \times
\,\frac{1}{a_1}\sum_{m_1 = -\infty}^{+\infty}\cdots\frac{1}{a_d}\sum_{m_d = -\infty}^{+\infty}\frac{k(\m)}{2}\biggl[e^{i[\k(\m)\cdot(\x-\x') - k(\m)(t-t')]}\theta(t-t') \nonumber \\ && +  e^{-i[\k(\m)\cdot(\x-\x') - k(\m)(t-t')]}\theta(t'-t)\biggr],
\end{eqnarray}
To obtain the respective correction to $G^{(1)}(x,x')$, one must employ the relation (\ref{apg2}) and notice that $G^{R,A}$ have similar perturbation expansions as $G$, Eq. (\ref{a1}). In this way, using the appropriate contour for $G^{R,A}$, one gets
\begin{equation}
\overline{\nu(x)\,G^{(1)}_{1}(x'',x')} = 2\biggl\{\overline{\nu(x)\,G_{1}(x'',x')} + \frac{1}{2}\Bigl[\,\overline{\nu(x)\,G^{R}_{1}(x'',x')} +\overline{\nu(x)\,G^{A}_{1}(x'',x')}\,\Bigr]\biggr\},
\label{h1mp}
\end{equation}
where
\begin{eqnarray}
\overline{\nu(x)\,G^{R}_{1}(x'',x')} &=& -\frac{\sigma^2}{a_1}\sum_{n_1 = -\infty}^{+\infty}\cdots\frac{1}{a_d}\sum_{n_d = -\infty}^{+\infty}\frac{1}{2k(\n)}\biggl[e^{i[\k(\n)\cdot(\x''-\x) - k(\n)(t''-t)]}  \nonumber \\ && - e^{-i[\k(\n)\cdot(\x''-\x) - k(\n)(t''-t)]}\biggr]\nonumber\\ && \times
\frac{1}{a_1}\sum_{m_1 = -\infty}^{+\infty}\cdots\frac{1}{a_d}\sum_{m_d = -\infty}^{+\infty}\frac{k(\m)}{2}\biggl[e^{i[\k(\m)\cdot(\x-\x') - k(\m)(t-t')]}\nonumber \\ && -  e^{-i[\k(\m)\cdot(\x-\x') - k(\m)(t-t')]}\biggr]\theta(t''-t)\theta(t-t'),
\end{eqnarray}
and
\begin{eqnarray}
\overline{\nu(x)\,G^{A}_{1}(x'',x')} &=&  -\frac{\sigma^2}{a_1}\sum_{n_1 = -\infty}^{+\infty}\cdots\frac{1}{a_d}\sum_{n_d = -\infty}^{+\infty}\frac{1}{2k(\n)}\biggl[e^{i[\k(\n)\cdot(\x''-\x) - k(\n)(t''-t)]}  \nonumber \\ && - e^{-i[\k(\n)\cdot(\x''-\x) - k(\n)(t''-t)]}\biggr]\nonumber\\ && \times
\frac{1}{a_1}\sum_{m_1 = -\infty}^{+\infty}\cdots\frac{1}{a_d}\sum_{m_d = -\infty}^{+\infty}\frac{k(\m)}{2}\biggl[e^{i[\k(\m)\cdot(\x-\x') - k(\m)(t-t')]}\nonumber \\ && -  e^{-i[\k(\m)\cdot(\x-\x') - k(\m)(t-t')]}\biggr]\theta(t-t'')\theta(t'-t).
\end{eqnarray}


\begin{thebibliography}{99}
%
\bibitem{carlip} S. Carlip, Rep. Prog. Phys. {\bf 64}, 885 (2001).
%
\bibitem{ash} A. Ashtekar, New. Jour. Phys. {\bf 7}, 198 (2005).
%
\bibitem{ford11} L. H. Ford, Phys. Rev. D {\bf 51}, 1692 (1995).
%
\bibitem{fn2} L. H. Ford and N. F. Svaiter, Phys. Rev. {\bf D54}, 2640 (1996),
L. H. Ford and N. F. Svaiter, Phys. Rev. {\bf D56}, 2226 (1997),
H. Yu and L. H. Ford,  Phys. Rev. D {\bf 60}, 084023 (1999),
R. T. Thompson and L. H. Ford,  Phys. Rev. D {\bf 78},
024014 (2008), R. T. Thompson and L. H. Ford, Class. Quant. Grav.
{\bf 25}, 154006 (2008), H. Yu, N. F. Svaiter and L. H. Ford, Phys. Rev. D {\bf 80},
124019 (2009).
%
\bibitem{casimir} H. B.  G. Casimir,
Proc. Kon. Ned. Akad. Wekf. {\bf 51}, 793 (1948).
%
\bibitem{plunien} G. Plunien, B. M\"uller and W. Greiner, Phys. Rep. {\bf 134},
87 (1986).
%
\bibitem{mamayev} A. A. Grib, S. G. Mamayev and V. M. Mostepanenko,
{\em Vacuum Quantum Effects in Strong Fields},
(Friedman Laboratory Publishing, St. Petesburg, 1994).
%
\bibitem{bordag} M. Bordag, U. Mohideen and V. M. Mostepanenko, Phys. Rep.
{\bf 353}, 1 (2001).
%
\bibitem{milton} K. A. Milton, {\em{The Casimir Effect:  Physical
Manifestation of Zero-Point Energy}} (World Scientific, Singapore, 2001).
%
\bibitem{milton2} K. A. Milton. J. Phys. A {\bf 37}, R209 (2004).
%
\bibitem{power} E. Arias, G. Krein, G. Menezes and N. F. Svaiter, ``Thermal Radiation from a Fluctuating Event Horizon", Int. J. Mod. Phys. A (to be published) [arXiv:1109.6080[hep-th]].
%
\bibitem{krein1} G. Krein, G. Menezes and N. F. Svaiter, Phys. Rev. Lett. {\bf 105}, 131301
(2010).
%
\bibitem{pe13} A. Ishimaru, {\em Wave Propagation
and Scattering in Random Media} (Academic, New York, 1978).
%
\bibitem{pe14} P. Sheng, {\em{Scattering and Localization of Classical
waves in Random Media}} (World Scientific, Singapure, 1990).
%
\bibitem{prd} E. Arias, E. Goulart, G. Krein, G. Menezes and N. F. Svaiter, Phys. Rev. D {\bf 83}, 125022 (2011).
%
\bibitem{p2} A. Edery, J. Math. Phys. {\bf 44}, 599 (2003).
%
\bibitem{ford80} L. Ford and N. F. Svaiter,  Phys. Rev. D {\bf 80}, 065034 (2009).
%
\bibitem{p3} L. H. Ford and N. F. Svaiter, J. Phys. Conf. Ser. {\bf 161}, 012034 (2009).
%
\bibitem{green}  L. H. Brown and G. J. Maclay, Phys. Rev. {\bf 184}, 1272
(1969), C. M. Bender and P.Hays, Phys. Rev. D {\bf 14}, 2622 (1976), D. Deutsch and P. Candelas, Phys. Rev. D {\bf 20},
3063 (1979), P. Candelas, Phys. Rev. D {\bf 21}, 2185 (1980).
%
\bibitem{sinal} J. R. Ruggiero, A. H. Zimerman and A. Villani, Rev. Bras. Fis. {\bf 7},
663 (1977); J. R. Ruggiero, A. Villani and A. H. Zimerman, J. Phys. A {\bf 13}, 767 (1980).
%
\bibitem{blau} S. K. Blau, M. Visser and A. Wipf, Nucl. Phys. B {\bf 310}, 163 (1988).
%
\bibitem{nami1} N. F. Svaiter and B. F. Svaiter, Jour. Phys. A {\bf 25}, 979 (1992);
B. F. Svaiter and N. F. Svaiter, Phys. Rev. D {\bf 47}, 4581 (1993); J. Math. Phys. {\bf 35}, 1840 (1994).
%
\bibitem{hawking} S. W. Hawking, Comm. Math. Phys. {\bf 55}, 133 (1977),
A. Voros, Comm. Math. Phys. {\bf 110}, 439 (1987).
%
\bibitem{zeta1} E. Elizalde, S. D. Odintsov, A. Romeo, A. A.
Bytsenko and S. Zerbini, {\em{Zeta Regularization Techniques and
Applications}} (World Scientific, Singapure, 1994).
%
\bibitem{zeta2} K. Kirsten, {\em{Spectral Functions in Mathematics and Physics}}
(Chapman and Hall/CRC, Florida, 2002).
%
\bibitem{fulling} S. Fulling, J. Phys. A {\bf 36}, 6857 (2003).
%
\bibitem{fluct} L. H. Ford and N. F. Svaiter,  Phys. Rev. D {\bf 58}, 065007-1 (1998).
%
\bibitem{davis} N. D. Birrell and P. C. Davis, {\em Quantum Fields in
Curved Space} (Cambridge University Press, New York, 1982).
%
\bibitem{time} M. J. Stephen, Phys. Rev. B {\bf 37}, 1 (1988).
%
\bibitem{ambjorn} J. Ambjorn and S. Wolfram, Ann. Phys. (NY) {\bf
147}, 1 (1983).
%
\bibitem{namis} F. Caruso, N. P. Neto,  B. F. Svaiter and N. F. Svaiter,
 Phys. Rev. D {\bf 43}, 1300 (1991),
R. D. M. De Paola, R. B. Rodrigues and N. F. Svaiter, Mod. Phys.
Lett. A {\bf 34}, 2353 (1999), L. E. Oxman. N. F. Svaiter and R. L. P. G.
Amaral, Phys. Rev. D {\bf 72}, 125007 (2005).
%


\end{thebibliography}
\end{document}